\documentclass[prb,preprint]{revtex4}

\usepackage[dvips]{graphicx}
\usepackage{amsmath,bm}

\bmdefine{\boldq}{q}  
\bmdefine{\boldP}{P}

\makeatletter
\def\@dotsep{4.5}
\makeatletter

\begin{document}
\title{Nonlinear Energy Response of Glass Forming Materials}
\author{Fumitaka Tagawa}
 \email{f.tagawa@cmt.phys.kyushu-u.ac.jp}
\author{Takashi Odagaki}
 \affiliation{Department of Physics, Kyushu University, Fukuoka 812-8581, Japan.}
\date{\today}


\begin{abstract}
 A theory for the nonlinear energy response of a system subjected to a
 heat bath is developed when the temperature of the heat bath is
 modulated sinusoidally.
 The theory is applied to a model glass forming system,
 where the landscape is assumed to have 20 basins and transition rates
 between basins obey a power law distribution.
 It is shown that  the statistics of eigenvalues of the transition rate matrix,
 the glass transition temperature $T_g$, the Vogel-Fulcher temperature
 $T_0$ and the crossover temperature $T_x$ can be determined from the 1st- and
 2nd-order ac specific heats, which are defined as coefficients
 of the 1st- and 2nd-order energy responses.
 The imaginary part of the 1st-order ac specific heat has a broad peak
 corresponding to
 the distribution of the eigenvalues.
 When the temperature is decreased below $T_g$,
 the frequency of the peak decreases and the width increases.
 Furthermore, the statistics of eigenvalues
 can be obtained from the frequency dependence of the 1st-order 
 ac specific heat.
 The 2nd-order ac specific heat shows extrema as a function of the
 frequency.
 The extrema diverge at the Vogel-Fulcher temperature $T_0$.
 The temperature dependence of the extrema
 changes significantly near $T_g$ and some extrema vanish near $T_x$.
\end{abstract}
\maketitle

\section{Introduction}

Since the anomaly of the specific heat at the glass transition
was discovered in 1923\cite{simon}, many studies have been conducted to
understand the behavior. Since the anomaly depends on the measurement process,
the transition is now believed not to be understood in the framework of the
standard thermodynamics\cite{parisi}.

Recently the concept of the landscape has been paid much attention because of
possibility to explain the transition in non-equilibrium systems 
\cite{goldstein, sciortino}.
In particular, the free energy landscape (FEL) picture proposed by
Odagaki et al \cite{odagaki06, odagaki07} is considered to provide
the unified concept for understanding thermodynamic and dynamic
singularities of the glass transition.
In fact, the single particle dynamics \cite{TDM} and
the specific
heat\cite{Odagaki, Odagaki2, Tao2, Tao3, Bagchi, Bagchi2, Tagawa}
were shown to phenomenologically be well described by the framework based on
the FEL.
Namely, the dynamical transition is understood as the Gaussian-to-Non-Gaussian
transition\cite{TDM}, and the thermodynamic singularities
near $T_g$, including the cooling rate dependence of the specific heat,
is characterized as the Quenched-to-Annealed transition in dynamics on the FEL
\cite{Odagaki, Odagaki2, Tao2, Tao3}.

The specific heat of non-equilibrium systems such as glass
forming materials is defined by the
reponse of the energy to the unit rise of temperature\cite{yamamuro}.
Because of the slow relaxations, the system cannot reach equilibrium
during the measurement and the energy response shows the time delay.
Therefore one can expect that the ac specific heat
\cite{Birge, Christ} will contain informations of the slow dynamics.
The ac specific heat $\tilde{C}_1(i\omega)$
is defined as the Laplace-Fourier transform of the energy or enthalpy
correlation function $\phi(t)$ \cite{Nielsen, Tagawa}
\begin{align}
 \tilde{C}_1(i\omega) = \int_0^{\infty} dt \phi(t) e^{-i\omega t}.
\end{align}
The ac specific heat is often fitted by the Laplace-Fourier transform
of the stretched exponential $\phi(t)=\exp\Big(-(t/\tau)^\beta\Big)$,
since the correlation function is believed to be the superposition of
the Debye relaxation function with different relaxation times.
However the origin of the distribution of the relaxation times has not
clearly been understood.

So far, studies of the energy response of glass forming systems
have been limited within the linear response region, and
the nonlinear energy response has not been studied yet.
In a previous paper\cite{Tagawa}, we proposed the description of the
1st- and 2nd-order energy response with the free energy landscape
picture and applied it to a two level system with a diverging barrier.
It is shown that the 2nd-order energy response has a
diverging term at the temperature where
the relaxation time diverges.

In this paper, we investigate the linear and nonlinear
energy responses of non-equilibrium systems described by the FEL
picture to an oscillating temperature and present the characteristic
behavior of the 1st- and 2nd-order ac specific heat.
Using a model FEL which supports a glass transition, 
we show that the Vogel-Fulcher temperature and the cross-over temperature
as well as the glass transition temperature can be determined from the
characteristic behavior of the ac specific heats.
We also show that the statistics of the transition rate matrix
representing the stochastic dynamics among the basins of the FEL
can be obtained from the frequency dependence of the 1st-order
ac specific heat.

The organization of the paper is as follows:
In \S 2, we explain the 1st- and 2nd-order energy responses
when the temperature of the heat bath is oscillated sinusoidally.
The definition of the 1st- and 2nd-order ac specific heat is also given.
In \S 3,
the stochastic dynamics on the free energy landscape is explained.
In \S 4,
we describe the 1st- and 2nd-order energy responses to the
oscillating temperature and the 1st- and 2nd-order ac specific heats
when the system is described by the FEL picture.
As an example, a model system with FEL consisting of 20 basins
is analyzed.
In \S 5, where transition rates between basins are assumed
to obey a power law distribution.
We present the characteristics of the 1st- and 2nd-order ac specific heats
of the glass former and show that the glass transition temperature,
the Vogel-Fulcher temperature and the crossover temperature
can be determined by the ac specific heats.
In \S 6, our conclusion is given.

\section{The first and second order specific heats}

The specific heat at the constant volume $C_V$
is conventionally defined by a derivative of the energy 
with respect to the temperature,
\begin{align}
 C_V(T) =\Big(  \frac{\partial E_{\rm eq}}{\partial T}  \Big)_V,
\label{cvoft}
\end{align}
where $E_{\rm eq}$ is the energy in the equilibrium state,
$T$ is the temperature and $V$ is the volume.
In this discussion of the specific heat,
it is not considered how long the system takes to equilibrate itself
when the temperature is changed.
When the system contains degrees of freedom of slow dynamics,
one must consider the effect of the delay in response.
We consider the energy response of a system with slow relaxations which is
subjected to a heat bath whose temperature is oscillated as $T+\Delta T(t)$,
where $T$ is the average temperature and $\Delta T(t)$ is the
oscillating part.
We assume that the energy response $\Delta E(t)$ can be expanded as follows:
\begin{align}
 \Delta& E(t) = \int_{-\infty}^{t}dt_1 C_1(t-t_1)\Delta T(t_1) \notag\\
 &+ \int_{-\infty}^{t}dt_1 \int_{-\infty}^{t}dt_2
 C_2(t-t_1,t-t_2)\Delta T(t_1)\Delta T(t_2)+O(\Delta T^3),
\label{eq:response_E}
\end{align}
where $C_1$ and $C_2$ represent the retardation effect of the system.

The Fourier transform $\Delta E(\omega)$
of Eq. (\ref{eq:response_E}) is given by
\begin{align}
 \Delta E(\omega) &= \tilde{C}_1(i\omega)\Delta T(\omega) \notag\\
 &+ \int_{-\infty}^{\infty} d\omega_1 
 \tilde{C}_2(i\omega_1,i\omega-i\omega_1)
 \Delta T(\omega_1)\Delta T(\omega-\omega_1)+O(\Delta T^3)
 \label{eq:response_E_FT},
\end{align}
where the Laplace components $\tilde{C}_1(p)$ and $\tilde{C}_2(p_1,p_2)$ 
are defined by
\begin{align}
 \tilde{C}_1(p) &= \int_0^{\infty}dt C_1(t)e^{-pt},
 \label{eq:C_p1}\\
 \tilde{C}_2(p_1,p_2)
 &= \int_0^{\infty}dt_1 \int_0^{\infty}dt_2
 C_2(t_1,t_2)e^{-p_1t_1}e^{-p_2t_2},
 \label{eq:C_p2}
\end{align}
and $\Delta T(\omega)$ is the Fourier transform of $\Delta T(t)$.

We now discuss the energy response $\Delta E(t)$ when 
$\Delta T(t)$ is a sinusoidal function
$\Delta T(t)=T_a\sin(\omega t)$,
where $T_a$ is the amplitude of the oscillating temperature.
It is straightforward to show that
\begin{align}
 \Delta E(t) &=  T_a \{\tilde{C}'_1(i\omega)\sin(\omega t)+
 \tilde{C}''_1(i\omega)\cos(\omega t)\} \notag\\
 &-\frac{T_a^2}{2}\{\tilde{C}'_2(i\omega,i\omega)\cos(2\omega t)
 -\tilde{C}''_2(i\omega,i\omega)\sin(2\omega t)
 -\tilde{C}'_2(i\omega,-i\omega)\} + O(T_a^3).
\end{align}
Here $C_1(t)$ and $C_2(t_1,t_2)$ are assumed to be real 
and the notations $'$ and $''$ represent
the real and imaginary parts, respectively.
The coefficients of the 1st-order temperature term,
$\tilde{C}'_1(i\omega)$ and $\tilde{C}''_1(i\omega)$ are known as 
the real and imaginary parts of the (1st-order) ac
specific heat,
which was introduced by Birge and Nagel\cite{Birge}
and by Christensen\cite{Christ}.

The 2nd-order temperature term consists of the oscillating and nonoscillating
terms.
We define the 2nd-order ac specific heat
$\tilde{C}'_2(i\omega,i\omega)$ and 
$\tilde{C}''_2(i\omega,i\omega)$
by the coefficients of the oscillating term.

The Laplace-Fourier transforms $\tilde{C}_1(p)$ and $\tilde{C}_2(p_1,p_2)$ are
related to the temperature derivative of the energy.
When the change of the temperature is slower than 
that of the time scale of the retardation effects $C_1(t)$ and $C_2(t_1,t_2)$,
it is straightforward to obtain the following expressions,
\begin{align}
 \lim_{p\to 0}&\tilde{C}_1(p) =
 \int_{0}^{\infty}dt C_1(t) = \frac{\partial E_{eq}}{\partial T} 
 \label{eq:Limit_C_1p}\\
 \lim_{p_1\to 0}&\lim_{p_2\to 0} \tilde{C}_2(p_1,p_2) =
 \int_{0}^{\infty}dt_1 \int_{0}^{\infty}dt_2C_2(t_1,t_2)
 = \frac{1}{2}\frac{\partial^2 E_{eq}}{\partial T^2}.
 \label{eq:Limit_C_2p}
\end{align}

When the frequency is smaller than the inverse of the structural
relaxation time, the energy responses without delay and
$\Delta E(t)$ can be expressed as
\begin{align}
 \Delta E(t) = \frac{\partial E_{eq}}{\partial T}\Delta T(t)
 + \frac{1}{2}\frac{\partial^2 E_{eq}}{\partial T^2}\Delta T^2(t)
 + O(\Delta T^3).
\end{align}
In this limit, $\tilde{C}_1'(i\omega)$ becomes equal to the specific heat 
in the equilibrium state,
$C_V=\partial E_{eq}/\partial T$
and $\tilde{C}_1''(i\omega)$ vanishes.
It can also be confirmed that in this limit
$\tilde{C}_2(i\omega,i\omega)$ and $\tilde{C}_2(i\omega,-i\omega)$
corresponds to the temperature derivative of the specific heat.
These behaviors are consistent with Eqs. (\ref{eq:Limit_C_1p}) and 
(\ref{eq:Limit_C_2p}).

\section{Dynamics on the free energy landscape}

The free energy surface in the configurational space is defined by
the partial partition function which is given by the partial summation
of the phase space spanned by the fast microsopic motion
\cite{odagaki06, odagaki07},
and the slow dynamics is represented by the stochastic motion
on the free energy surface.
Around the glass transition point,
the stochastic motions can be classified into two types;
the fluctuation in one basin due to the structural fluctuation around
a certain structure and 
the transition between basins which corresponds to the structural
relaxation.

Here, we concentrate on the transition between basins
and ignore the fluctuation within a single basin.
We denote the probability that the system is in basin $i$ at
time $t$ at temperature $T$ by $P_i(T,t)$
and a physical quantity $A$ of basin $i$ by $A_i(T)$.
Then the physical quantity we measure at time $t$ 
is defined as an average over the basins,
\begin{align}
 A(T, t)=\sum_i A_i(T)P_i(T,t) = \vec{A}(T)\cdot\vec{P}(T,t),
\label{property}
\end{align}
where $\vec{A}(T)$ and $\vec{P}(T,t)$ are the vectors 
consisting of components $A_i(T)$ and $P_i(T,t)$, respectively.

The probability vector $\vec{P}(T,t)$ is assumed to obey the master
equation,
\begin{align}
 \frac{d}{dt}\vec{P}(T,t) = W(T)\vec{P}(T,t), \label{eq:master}
\end{align}
where $W(T)$ is the transition rate matrix, 
i.e. $W_{ij}$ is the transition rate from basin $j$ to $i$ and 
$W_{ii}=-\sum_{j\neq i}W_{ji}$ is the transition rate jumping out 
from basin $i$. The transition rate is assumed to be related to the
free energy barrier as follows,
\begin{align}
 W_{ij} = C\exp[-\beta(F_A(T)-F_j(T))],
\end{align}
where $C$ is the attempt frequency of the jump motion,
$\beta=1/k_B T$ is the inverse of the temperature $T$ multiplied by
Boltzmann constant $k_B$,
$F_i(T)$ is the free energy of basin $i$,
$F_A(T)=\max\{F_i(T),F_j(T)\}+\Delta_{i,j}(T)$
and $\Delta_{i,j}(T)$ is the energy barrier between basin $i$ and $j$.
In the high temperature region, the transition rate is large
and the system moves among basins freely.
On the other hand, the transition rate becomes small
and the structural transition is hindered at low temperatures.

The transition rate matrix $W$ in the master equation (\ref{eq:master})
must satisfy the condition that the long time limit of $P_i(T,t)$ becomes the
Boltzmann distribution
\begin{align}
\lim_{t\to\infty} P_i(T,t) =
P_{i}^{eq}(T)=\frac{\exp[-\beta F_i(T)]}{\sum_j\exp[-\beta F_j(T)]}.
\end{align}
Therefore, the transition rate matrix $W$ obeys the following detailed balance
\begin{align}
 W\vec{P}^{eq}=0 .\label{eq:detail}
\end{align}
When the temperature $T$ does not depend on time, 
Eq. (\ref{eq:master}) can be readily solved
\begin{align}
 \vec{P}(T,t) &= \exp[W(T)t]\vec{P}(T,t=0)\notag\\
 &= V\begin{pmatrix}
      \exp(\lambda_1t) & & &  \\
      & \ddots \\
       & & \exp(\lambda_{N-1}t)& \\
       & & & \exp(\lambda_N t)
     \end{pmatrix}
 V^{-1}
 \vec{P}(T,t=0),\label{eq:master_kai}
\end{align}
where $N$ is the number of basins,
$\lambda_i$ is the $i$-th eigenvalue of $W$ 
and $V$ is a matrix whose columns
are eigenvectors of $W$.
It is important to note that
there is an eigenvalue $\lambda_i =0 $
which corresponds to the detailed balance of Eq. (\ref{eq:detail}). 

\section{The energy response to the oscillating temperature}
\label{sec:Theory}
\subsection{The 1st- and 2nd-order energy response}

The time dependent temperature $\hat{T}(t)$ is assumed to be
$\hat{T}(t) = T+ \Delta T(t)$,
where $T$ is the average temperature and $\Delta T(t)$ is the
oscillating part of the temperature.
The probability vector $\vec{P}(T, t)$ is expanded as
\begin{align}
 \vec{P}(\hat{T}(t),t) &= \vec{P}_0(t)
 + \Delta \vec{P}_1(t) + \Delta \vec{P}_{2}(t) + O(\Delta T^3),
 \label{eq:p_expansion}
\end{align}
where $\Delta\vec{P}_n$ is the term of order $\Delta T^n$.
Note the explicit temerature dependence of the quantities
on the right hand side of Eq. (\ref{eq:p_expansion}).
is omitted.
It is clear that $\vec{P}_0(t)=\vec{P}^{eq}(T)$.
Then the 1st- and 2nd-order energy responses,
$\Delta E_1(t)$ and $\Delta E_2(t)$ at time $t$ are represented as
\begin{align}
 \Delta E_1(t) &= \vec{E}\cdot\Delta \vec{P}_1(t) + 
  \frac{\partial \vec{E}}{\partial T}\cdot\vec{P}_{eq}(T)
  \Delta T(t) \label{eq:FELP_E1}\\
 \Delta E_2(t) &=\vec{E}\cdot\Delta\vec{P}_2(t)
 + \frac{\partial \vec{E}}{\partial T}\Delta\vec{P}_{1}(t)\Delta T(t)
 +\frac{1}{2}\frac{\partial^2 \vec{E}}{\partial
 T^2} \vec{P}_{eq}(T)\Delta T^2(t) \label{eq:FELP_E2}
\end{align}
where $\vec{E}$ is the vector consisting of components $E_i$.
The 1st- and 2nd-order probability responses,
$\Delta P_1(t)$ and $\Delta P_2(t)$,
obey the following equations derived from Eq. (\ref{eq:master}),
\begin{align}
 \frac{d}{dt}\Delta\vec{P_1}(t) &= W(T)\Delta\vec{P_1}(t)
 +\frac{\partial W(T)}{\partial T}\vec{P}_{eq}(T)\Delta T(t)
\label{eq:1ji_eq}\\
 \frac{d}{dt}\Delta\vec{P_2}(t) &= W(T)\Delta\vec{P_2}(t)
 +\frac{\partial W(T)}{\partial T}\Delta\vec{P_1}(t)\Delta T(t) 
 +\frac{1}{2}\frac{\partial^2 W(T)}{\partial T^2}\vec{P_{eq}}(T)\Delta
 T(t)^2.
 \label{eq:2ji_eq}
\end{align}
It is straightforward to solve Eqs. (\ref{eq:1ji_eq}) and (\ref{eq:2ji_eq}) as
\begin{align}
 \Delta\vec{P}_1(t) &= \int_{-\infty}^t ds e^{W(t-s)}
 \frac{\partial W}{\partial T}\vec{P}_{eq}(T)\Delta T(s), \label{eq:1ji_kai}\\
 \Delta\vec{P}_2(t) &= \int_{-\infty}^t ds_1 \int_{-\infty}^{s_1} ds_2
 e^{W(t-s_1)}\frac{\partial W}{\partial T}e^{W(s_1-s_2)}
 \frac{\partial W}{\partial T}\vec{P}_{eq}(T)\Delta T(s_1)\Delta T(s_2) \notag \\
 &+\frac{1}{2}\int_{-\infty}^t ds e^{W(t-s)}
 \frac{\partial^2 W}{\partial T^2}\Delta T^2(s) \label{eq:2ji_kai}.
\end{align}

\subsection{The 1st-order energy response and the ac specific heat}
\label{sec:1st-order-ac}
The (1st-order) ac specific heat is defined as the linear coefficient of the
energy response to the temperature change\cite{Bagchi}.
Comparing Eqs. (\ref{eq:response_E}), (\ref{eq:FELP_E1}) 
and (\ref{eq:1ji_kai}),
the retardation effect of the 1st-order energy response
$C_1(t)$ is represented in the free energy landscape picture as follows,
\begin{align}
 C&_1(t) =
 2C_0(T)\delta(t)
 +\vec{E}\cdot\exp(Wt)\frac{\partial W}{\partial T}\vec{P}_{eq},
 \label{eq:FELP_C1t}
\end{align}
where $C_0(T)=\vec{P}_{eq}\cdot\partial \vec{E}/\partial T$ is 
the quenched specific heat contributed only from the fast degree of freedom
at one basin\cite{Odagaki, Odagaki2, Tao2, Tao3, Tagawa}.
The 2nd term represents the response from
the transition between basins.

\begin{figure} 
 \begin{center}
  \includegraphics[width=8cm]{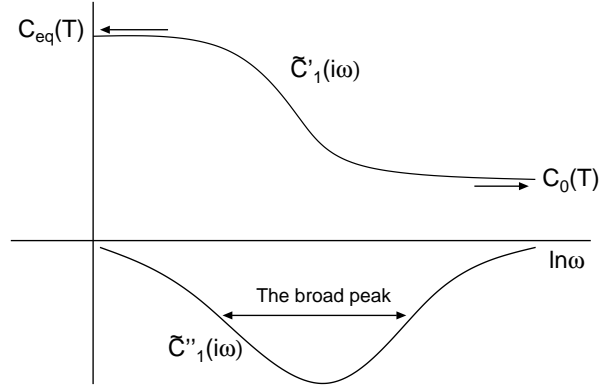}
 \end{center}
 \caption{The schematic picture of the (1st-order) ac specific heat 
 in the free energy landscape picture.}
 \label{fig:ac_specific_heat_FELP}
\end{figure}

From Eq.(\ref{eq:C_p1}), the 1st-order ac specific heat $\tilde{C}_1(i\omega)$
is given by the Laplace-Fourier transform of $C_1(t)$.
The real part $\tilde{C}'_1(i\omega)$ and 
the imaginary part $\tilde{C}''_1(i\omega)$ are written as
\begin{align}
 \tilde{C}'_1(i\omega)
 &= C_{0}(T)
 +\vec{E}\cdot V
  \begin{pmatrix}
   \lambda_1^2/(\omega^2 + \lambda_1^2)\\
   &\ddots\\
   & & \lambda_N^2/(\omega^2 + \lambda_N^2)
  \end{pmatrix}
 V^{-1}\frac{\partial \vec{P}_{eq}}{\partial T}, 
 \label{eq:AC1_Real}\\
 \tilde{C}''_1(i\omega)
 &= \vec{E}\cdot V
  \begin{pmatrix}
   \lambda_1\omega/(\omega^2 + \lambda_1^2)\\
   &\ddots\\
   & &  \lambda_N\omega/(\omega^2 + \lambda_N^2)
  \end{pmatrix}
 V^{-1}\frac{\partial \vec{P}_{eq}}{\partial T}.
 \label{eq:AC1_Imag}
\end{align}

The frequency dependence of the ac specific heat is shown 
in Fig.\ref{fig:ac_specific_heat_FELP} schematically.
In the low frequency region, $\tilde{C}'_1(i\omega)$ approaches
the specific heat in the equilibrium state as
$C_{eq}(T) = \partial\vec{E}/\partial T\cdot\vec{P}^{eq}
+ \vec{E}\cdot\vec\partial\vec{P}^{eq}/\partial T$ 
and $\tilde{C}''_1(i\omega)$ vanishes.
This indicates the fact that the energy responds without time delay 
when the temperature is oscillated slowly.
Since the eigenvalue $\lambda_i$ is distributed, 
$\tilde{C}''_1(i\omega)$ has a broader peak than that
of the Debye relaxation type and
$\tilde{C}'_1(i\omega)$
is represented as the sum of Lorentzian funtions.
This behavior is consistent with the measurement of the ac specific heat
in glass forming materials\cite{Birge}.
In the high frequency region, $\tilde{C}'_1(i\omega)$ approaches $C_{0}(T)$
and $\tilde{C}''_1(i\omega)$ vanishes.
It exhibits that the energy response within the 1st-order temperature
perturbation is determined by the energy change in a basin alone
when the temperature fluctuates rapidly.

The 1st-order ac specific heat can be used as the method to obtain the
statistics of eigenvalues $\lambda_i$.
From Eqs.(\ref{eq:AC1_Real}) and (\ref{eq:AC1_Imag}),
the scaled 1st-order ac specific heat is related to
the distribution of the eigenvalues $\lambda_i$ 
of the transition rate matrix as follows,
\begin{align}
 &\frac{\tilde{C}'_1(i\omega)-C_{0}}{C_{eq}-C_{0}}
 = \sum_ig_i\frac{\lambda_i^2}{\lambda_i^2+\omega^2}, \\
 &\frac{\tilde{C}''_1(i\omega)}{C_{eq}-C_{0}}
 = \sum_ig_i\frac{\lambda_i\omega}{\omega^2+\lambda_i^2}.
\end{align}
Here, $g_i$ is a factor related to the vectors $\vec{E}$, $\vec{P}_{eq}$
and $V$ given by 
\begin{align}
 g_i=\frac{\sum_{j,k}E_iV_{ji}V^{-1}_{ik}\frac{\partial P_{eq}^k}{\partial T}}
 {C_{eq}-C_{0}}.
\end{align}
In the small frequency region, the imaginary part behaves
as 
$\lim_{\omega\to 0}\tilde{C}_1"(i\omega)/(C_{eq}-C_0)=\sum_ig_i\lambda_i^{-1}\omega$.In the high frequency region, the real part behaves as 
$\lim_{\omega\to\infty}(\tilde{C}_1'(i\omega)-C_0)/(C_{eq}-C_0)=\sum_ig_i\lambda_i^2/\omega^2$
and the imaginary part behaves as
$\lim_{\omega\to\infty}\tilde{C}_1"(i\omega)/(C_{eq}-C_0)=\sum_ig_i\lambda_i\omega^{-1}$.
We can estimate $\sum_ig_i\lambda_i$, $\sum_ig_i\lambda_i^2$ 
and $\sum_ig_i\lambda_i^{-1}$
from the frequency dependence of the scaled 1st-order ac specific heat.
These properties correspond
to $\langle\lambda\rangle$, $\langle\lambda^2\rangle$ 
and $\langle\lambda^{-1}\rangle$ respectively as will be shown in Sec. 
\ref{sec:Model}.

\subsection{The 2nd-order ac specific heat}

The retardation effect of the 2nd-order energy response $C_2(t_1,t_2)$ is
represented from Eqs. (\ref{eq:response_E}), (\ref{eq:FELP_E2})
 and (\ref{eq:2ji_kai}) as
\begin{align}
 C&_2(t_1,t_2) = 
 2\frac{\partial C_{0}}{\partial T}\delta(t_1)\delta(t_2)+
 \vec{E}\cdot\exp(Wt_1)\frac{\partial W}{\partial T}\exp(Wt_2)
 \frac{\partial W}{\partial T}\vec{P}^{eq}\theta(t_2-t_1)\notag\\
 &+\frac{1}{2}\vec{E}\cdot\exp(Wt_1)\frac{\partial^2 W}{\partial T^2}
 \vec{P}^{eq}\delta(t_1-t_2)
 +2\frac{\partial \vec{E}}{\partial T}\exp(Wt_1)
 \frac{\partial W}{\partial T}\vec{P}^{eq}\delta(t_2).
\end{align}
The 1st term expresses the response due to 
the instant change of the specific heat of each basin
and the remaining terms correspond to 
the effect of the transition motion between basins.
Here, $\delta (x)$ is the Dirac's delta function 
and $\theta (x)$ is the Heaviside step function.

From the definition (\ref{eq:C_p2}),
the Laplace transform $\tilde{C}_2(p_1,p_2)$ is represented as
\begin{align}
\tilde{C}_2(p_1,p_2) 
 &=\frac{1}{2}\frac{\partial C_{eq}}{\partial T}-p_2\vec{E}\cdot((p_1+p_2)\hat{1}-W)^{-1}
 \frac{\partial W}{\partial T}(p_2\hat{1}-W)^{-1}
 \frac{\partial \vec{P}^{eq}}{\partial T}\notag\\
 &-\frac{p_1+p_2}{2}\vec{E}\cdot((p_1+p_2)\hat{1}-W)^{-1}\frac{\partial^2
 \vec{P}^{eq}}{\partial T^2}-p_1\frac{\partial \vec{E}}{\partial T}(p_1\hat{1}-W)^{-1}
 \frac{\partial \vec{P}^{eq}}{\partial T}
 \label{eq:FELP_C2pp}
\end{align}

The 2nd-order ac specific heats, which are introduced in
eq. (\ref{eq:response_E}) by the coefficients of the oscillating
term in the 2nd-order energy response, 
$\tilde{C}_2(i\omega,i\omega)=\tilde{C}'_2(i\omega,i\omega)
+i\tilde{C}''_2(i\omega,i\omega)$
are expressed as
\begin{align}
 \tilde{C}'_2&(i\omega,i\omega) = 
 \frac{1}{2}\frac{\partial C_{eq}}{\partial T}
 -2\omega^2\vec{E}\cdot(W^2+4\omega^2)^{-1}
 \frac{\partial^2\vec{P}^{eq}}{\partial T^2}\notag\\
 &+\omega^2\vec{E}\cdot(W^2+4\omega^2)^{-1}
 \Bigr\{
 2\frac{\partial W}{\partial T}W+W\frac{\partial W}{\partial T}
 \Bigl\}
 (W^2+\omega^2)^{-1}\frac{\partial \vec{P}^{eq}}{\partial T}\notag\\
 &-\omega^2\frac{\partial\vec{E}}{\partial T}(W^2+\omega^2)^{-1}
 \frac{\partial \vec{P}^{eq}}{\partial T},
 \label{eq:FELP_2nd_Real}
 \\
 \tilde{C}''_2&(i\omega,i\omega)= 
 \omega\vec{E}\cdot W(W^2+4\omega^2)^{-1}
 \frac{\partial^2 \vec{P}^{eq}}{\partial T^2} \notag\\
 &+\omega\vec{E}\cdot(W^2+4\omega^2)^{-1}
 \Bigr\{
 2\omega^2\frac{\partial W}{\partial T}-W\frac{\partial W}{\partial T}W
 \Bigl\}
 (W^2+\omega^2)^{-1}\frac{\partial \vec{P}^{eq}}{\partial T} \notag\\
 &+\omega\frac{\partial\vec{E}}{\partial
 T}(W^2+\omega^2)^{-1}\frac{\partial\vec{P}^{eq}}{\partial T}.
 \label{eq:FELP_2nd_Imag}
\end{align}
In the low frequency limit, $\tilde{C}'_2(i\omega,i\omega)$
becomes $(\partial C_{eq}/\partial T)/2$ and
$\tilde{C}''_2(i\omega,i\omega)$ vanishes.
In the high frequency limit,
$\tilde{C}'_2(i\omega,i\omega)$ is equal to 
$(\partial C_{0}/\partial T)/2$
and $\tilde{C}''_2(i\omega,i\omega)$ vanishes.
These behaviors are qualitatively similar to those of the 1st-order
ac specific heat.
It indicates that the structural change of the system can catch up with 
the temperature change in the small frequency region and
can not in the high frequency region.

Terms including the differential of the transition rate
with respect to the temperature, $\partial W/\partial T$,
 is expected to show the characteristic
behavior
in the low temperature region, where the transition matrix $W$ depends
strongly on the temperature change.
This characteristic does not appear in the 1st-order ac specific heat.
In the high temperature region, the temperature dependence of
the transition rate matrix $W$ is weak. Therefore, the terms including
$\partial W/\partial T$ are negligible,
and the 2nd-order ac specific heat becomes the superposition of the
susceptibility of the Debye relaxation.
We explain this behavior for a trap model in the next section.

The 2nd-order energy response
can be measured in the materials, 
where the derivatives of the quenched and equilibrium specific heats
are much different,
since the order of the 2nd-order energy response 
is the difference between derivatives of these
specific heats, $\partial (C_{eq}-C_{0})/\partial T$.

\section{Application To A Model Free Energy Landscape Of Glass Forming Systems}
\label{sec:Model}

\subsection{Model}

To clarify the characteristics of the 1st- and 2nd-order ac specific heat
of glass forming materials,
we apply the present analysis of the ac specific heat to
a model landscape incorporated with the distribution of the transition rate 
exploited in the trapping diffusion model \cite{TDM, Odagaki_PRL}.

We prepare a landscape consisting of 20 basins which are mutually connected.
The microscopic dynamics of the system in each basin is assumed to be the Debye oscillator. The free energy, $F_i$, of basin $i$ is given by
\begin{align}
 &F_i = \epsilon_i
 +  \exp\Big\{9 
 \Big(\frac{T}{\Theta_D}
 \Big)^3
 \int_0^{\Theta_D/T}
 \ln\Big(\frac{1}{2}\sin^{-1}\frac{x}{2} 
 \Big)x^2dx
 \Big\},
\end{align}
where 
the energy $\epsilon_i$ of the bottom of basin $i$ is uniformly
distributed with the variance $\epsilon$
and $\Theta_D$ is the Debye temperature.
Note that the form of the distribution does not play any important roles
in the following discussion.

The energy, $E_i$, of basin $i$ is calculated from the temperature
derivative of the free energy $F_i$.
The transition rate $w$ is assumed to obey the power law distribution
$P(w)$ as in the trapping diffusion model\cite{TDM, Odagaki_PRL},
\begin{align}
 P(w)= \begin{cases}
	\frac{\rho+1}{w_0}
	\Big(\frac{w}{w_0}\Big)^{\rho} & \text{$(0\leq w\leq w_0)$} \\
	0 & \text{otherwise}
       \end{cases},
\end{align}
where $\rho$ is related to the configurational entropy $s_c(T)$ as 
\begin{align}
 \rho = \frac{TS_c(T)-T_gS_c(T_g)}{T_gS_c(T_g)}.
\end{align}
The transition rate from basin $j$ to $i$ $W_{ij}$ is now a random
variable given by
\begin{align}
 W_{ij}&=w_0x^{\frac{1}{\rho+1}}\exp[-\beta(F_A(T)-F_j(T))].
 \label{eq:Trap_Transition_Rate}
\end{align}
Here, $x$ is the uniform random number in $[0,1]$.

For numerical calcuations in this study,
the exponent $\rho$ of the jump rate distribution is simplified as
\begin{align}
 \rho = \frac{T-T_g}{T_g-T_0},
\end{align}
i.e. $S_c(T)$ is assumed as $TS_c(T)\propto T-T_0$ with the
Vogel-Fulcher temperature $T_0\sim T_k$
and
the parameters are chosen as $T_0/T_g=0.75$
and $\Theta_D/T_g = 12.5$, respectively.
In addition, the crossover temperature $T_x$ is $T_x/T_g = 1.25$
since the crossover temperature is identified with the exponent 
$\rho = 1.0$ \cite{Odagaki_PRL}.

\subsection{The 1st-order ac specific heat}
\begin{figure}
 \begin{center}
  \includegraphics[width=8cm]{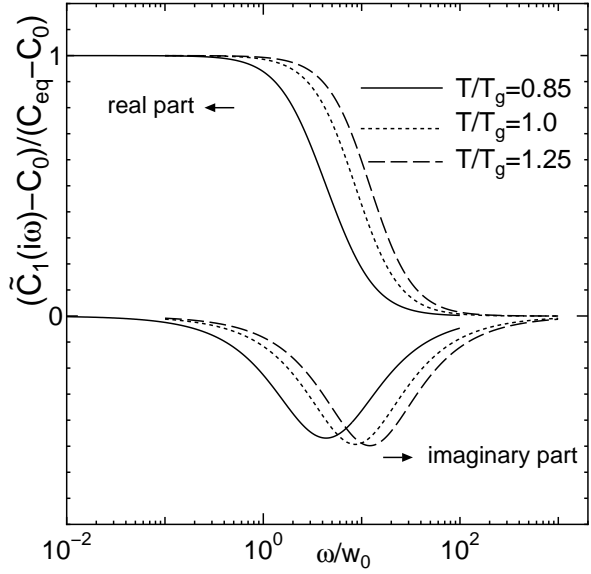}
  \end{center}
 \caption{The scaled ac specific heat in the trap model at the temperature 
 $T/T_g=0.85$ (solid line), $1.0$ (dot line), $1.25$ (dash line):
 }
 \label{fig:Trap_Model_AC1}
\end{figure}

The real and imaginary parts of the scaled 1st-order ac specific heat 
$(\tilde{C}_1(i\omega)-C_{0})/(C_{eq}-C_{0})$ are shown
in Fig.\ref{fig:Trap_Model_AC1}.
As the temperature is reduced,
a peak of the imaginary part shifts to the low frequency region and the
width of the peak increases.

\begin{figure}
 \begin{center}
  \includegraphics[width=8cm]{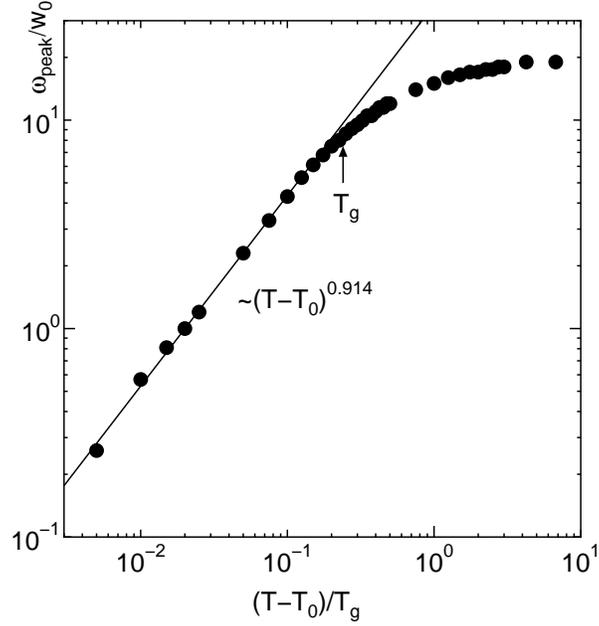}
 \end{center}
 \caption{The temperature dependence of the frequency at the peak
 of the imaginary part of the 1st-order ac specific heat:
 The peak frequency (black circle) becomes small as the temperature
 reduces toward the Vogel-Fulcher temperature $T_0$.
 The line is fitted with the data in the low temperature region.}
 \label{fig:Trap_Model_AC1_imag_peak}
\end{figure}

Figure \ref{fig:Trap_Model_AC1_imag_peak} shows the temperature
dependence of the frequency $\omega_{peak}$ 
at the peak of the imaginary part.
In the low temperature region below $T_g$,
$\omega_{peak}$ obeys the power law function
as $\omega_{peak}\sim(T-T_0)^{0.914}$.
In the high temperature region above $T_g$, 
the temperature dependence changes
and $\omega_{peak}$ approaches toward 
$20w_0$.
It is straightforward to show that,
when $N$ basins are mutually connected, the $\omega_{peak}$
becomes $Nw_0$ in the high temperature region.
In fact from Eq. (\ref{eq:Trap_Transition_Rate}), all transition rates
are $w_0$ when the temperature is much higher than
the glass transition temperature $T_g$.
The relaxation time due to eigenvalues
of transition matrix $W$ is not distributed
and the 1st-order ac specific heat becomes the same as the
susceptibility of the Debye relaxation.
The master equation (\ref{eq:master}) can be solved without difficulties 
in this limit as
$P_i(T,t)=e^{-Nw_0t}(P_i(t=0)-1/N)+1/N$.
Thus the peak frequency $\omega_{peak}$ becomes $Nw_0$ in
the high temperature region.

\begin{figure}
 \begin{center}
  \includegraphics[width=8cm]{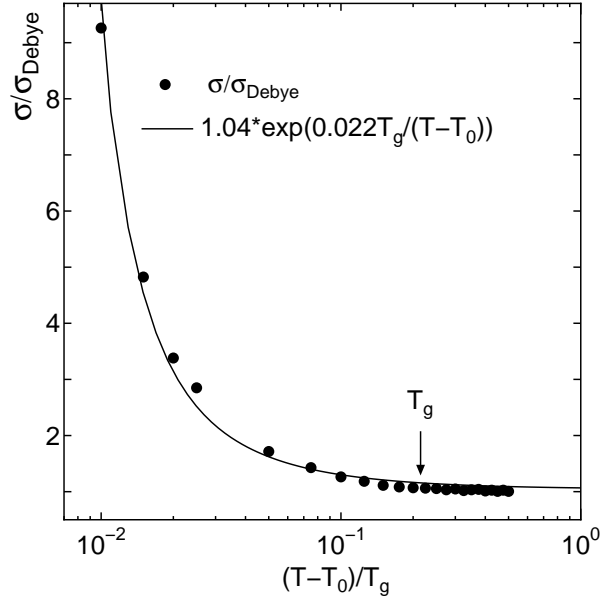}
 \end{center}
 \caption{The temperature dependence of the stretching factor $\sigma$
 of the imaginary part.
 The line represents the fitted line by Vogel-Fulcher law.
 Here, $\sigma_{Debye}=(2+\sqrt{3})/(2-\sqrt{3})$ 
 is the stretching factor for the Debye relaxation type.
 As the temperature is reduced toward the Vogel-Fulcher temperature $T_0$, 
 $\sigma$ increases.}
 \label{fig:Trap_Model_AC1_imag_width}
\end{figure}

We define the stretching factor $\sigma$ by the ratio 
$\sigma\equiv\omega_+/\omega_-(\omega_+>\omega_-)$.
Here, $\omega_\pm$ is the frequency satisfying the relation 
$\tilde{C}_1''(i\omega_\pm)=\tilde{C}_1''(i\omega_{peak})/2$.
Figure \ref{fig:Trap_Model_AC1_imag_width} shows the temperature
dependence of the stretching factor $\sigma$ as a function of temperature.
In the high temperature region above $T_g$, 
$\sigma$ is the same as that of the Debye susceptibility.
As the temperature is reduced toward $T_0$,
$\sigma$ of the peak increases and diverges at $T_0$.
It means that the relaxation time is distributed
and the broad peak appears in $\tilde{C}''_1(i\omega)$ below $T_g$.
It is interesting to note that $\sigma$ can be fitted well by the
Vogel-Fulcher law as $\sigma\sim\exp(A/(T-T_0))$ 
shown in Fig.\ref{fig:Trap_Model_AC1_imag_width}.

\begin{figure}
 \leavevmode
 \begin{center}
  \begin{tabular}{ c c c }
   \includegraphics[width=5cm]{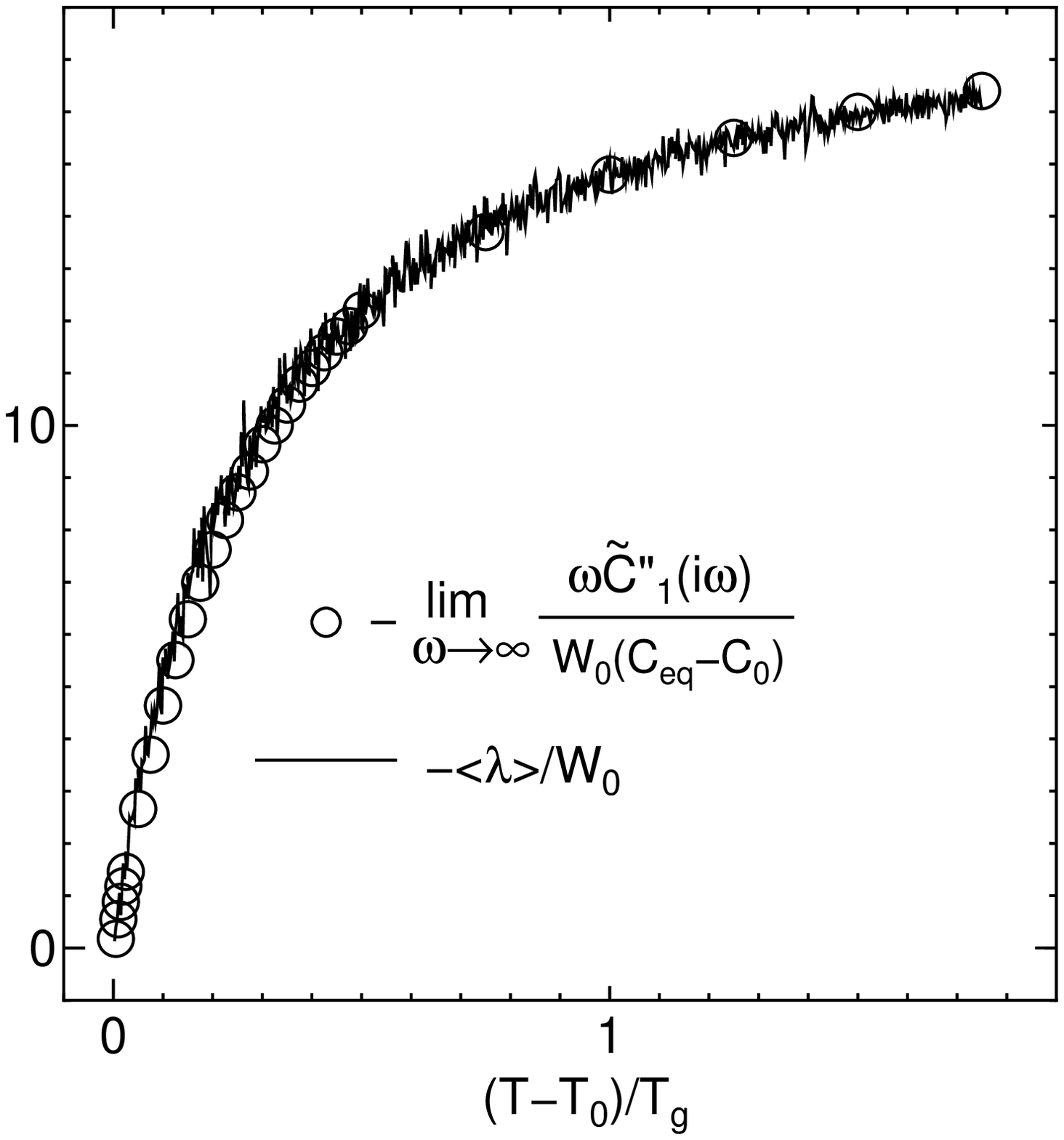}
   &
   \includegraphics[width=5cm]{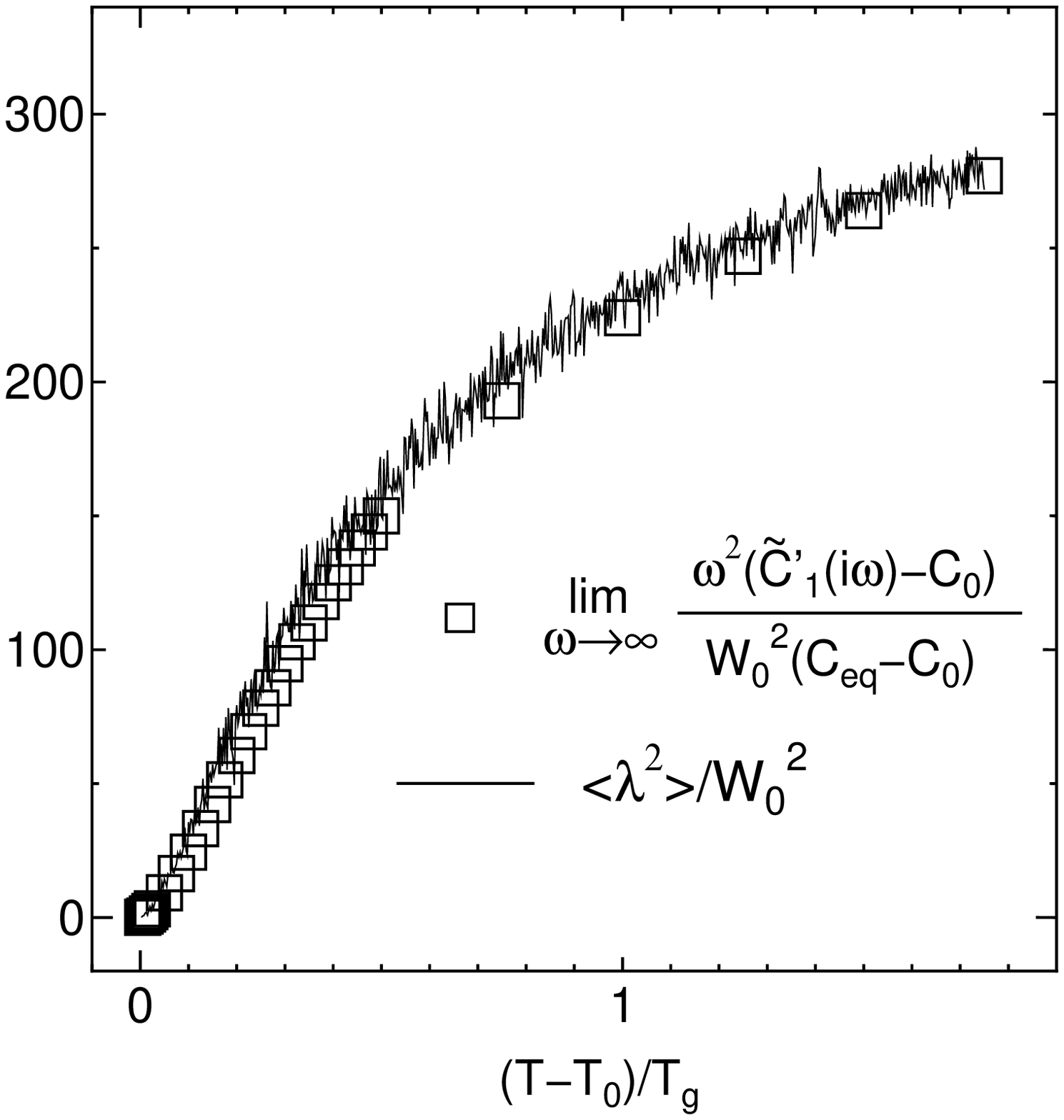}
   &
   \includegraphics[width=5cm]{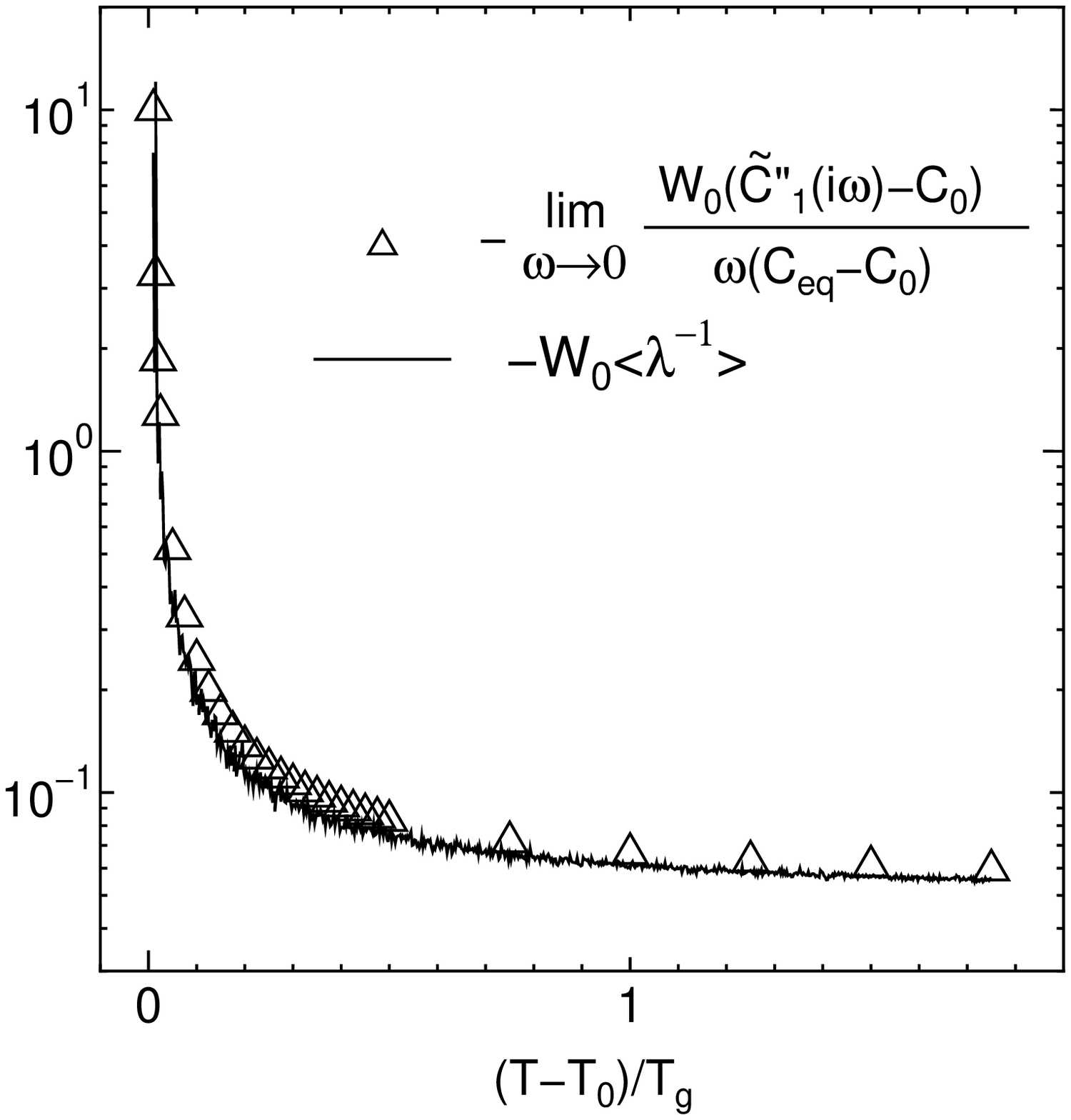}\\
   (a)
   &
   (b)
   &
   (c)
  \end{tabular}
 \caption{
 The temperature dependence of 
 (a)$-\lim_{\omega\to\infty}\omega\tilde{C}"_1(i\omega)
/(C_{eq}-C_0)/W_0$,
 (b)$\lim_{\omega\to\infty}\omega^2
 (\tilde{C}'_1(i\omega)-C_0)
/(C_{eq}-C_0)/W_0^2$,
 and (c)$-W_0\lim_{\omega\to 0}\omega^{-1}
 (\tilde{C}"_1(i\omega)-C_0)
/(C_{eq}-C_0)$
 determined from the frequency dependence of the 1st-order ac specific heat.
 The solid lines represent
 (a)$-\langle\lambda\rangle /W_0$, (b)$\langle\lambda^2\rangle /W_0^2$ 
 and (c)$-W_0\langle\lambda^{-1}\rangle$,
 which are calculated numerically from eigenvalues of
 the transition matrix $W$, respectively.
 } \label{fig:Trap_AC1_lambda}
 \end{center}
\end{figure}

The statistics of eigenvalues of the transition rate matrix $W$
can be obtained from the frequency dependence 
of $\tilde{C}_1(i\omega)$.(See \S\ref{sec:1st-order-ac}.)
We show a result of comparison of
$\lim_{\omega\to\infty}\omega\tilde{C}"_1(i\omega)
/(C_{eq}-C_0)$,
$\lim_{\omega\to\infty}\omega^2(\tilde{C}'_1(i\omega)-C_0)
/(C_{eq}-C_0)$,
and 
$\lim_{\omega\to 0}\omega^{-1}\tilde{C}"_1(i\omega)
/(C_{eq}-C_0)$
with statistics of eigenvalues 
$\langle \lambda \rangle$, $\langle \lambda^2 \rangle$
and 
$\langle\lambda^{-1}\rangle$
in Fig.\ref{fig:Trap_AC1_lambda}.
Here, 
$\langle f(\lambda)\rangle$ represents the statistics of eigenvalues as
$\langle f(\lambda) \rangle=\sum_{i=1,\lambda_i\neq 0}^{20}f(\lambda_i)/20$,
which is calculated numerically from the transition matrix 
$W$.
The result indicates that the contribution of $g_i$ is negligible
and the coefficients obtained from the frequency dependence of the
1st-order ac specific heat can be used to estimate
the statistics of eigenvalues of transition rate matrix.

\begin{figure}
 \begin{center}
   \includegraphics[width=8cm]{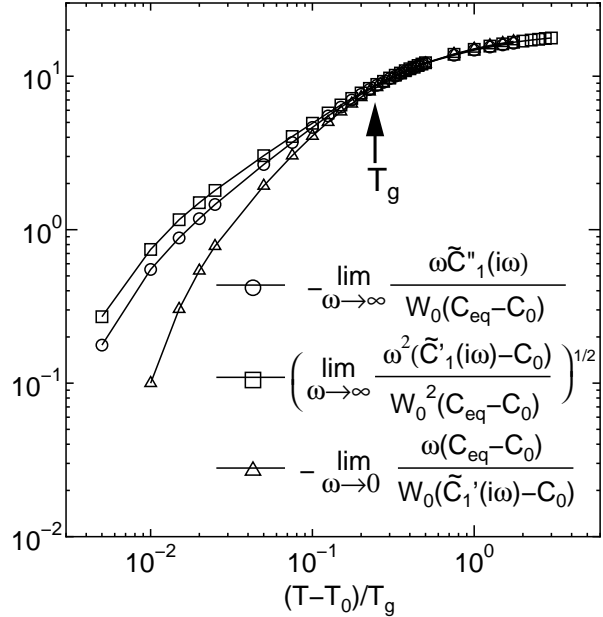}
 \end{center}
 \caption{The temperature dependence of 
 $-\lim_{\omega\to\infty}\omega\tilde{C}"_1(i\omega)
/(C_{eq}-C_0)/W_0$,
 $\lim_{\omega\to\infty}\{\omega^2
 (\tilde{C}'_1(i\omega)-C_0)
/(C_{eq}-C_0)/W_0^2\}^{1/2}$,
 and 
 $-\lim_{\omega\to 0}\omega
(C_{eq}-C_0)/(\tilde{C}'_1(i\omega)-C_0)/W_0$.
 The deviation of these properties becomes significant below $T_g$.}
 \label{fig:AC1_stat_all}
\end{figure}

Figure \ref{fig:AC1_stat_all} shows the temperature dependence of
$\lim_{\omega\to\infty}\omega\tilde{C}"_1(i\omega)
/(C_{eq}-C_0)/W_0$,
$\lim_{\omega\to\infty}\{\omega^2(\tilde{C}'_1(i\omega)-C_0)
/(C_{eq}-C_0)/W_0^2\}^{1/2}$,
and 
$\lim_{\omega\to 0}\omega(C_{eq}-C_0)/(\tilde{C}'_1(i\omega)-C_0)/W_0$
and the deviation between these statistics
becomes significant below $T_g$.
It reflects that the distribution of the eigenvalues becomes broader
below $T_g$.


\subsection{The 2nd-order ac specific heat}
\begin{figure}
 \begin{center}
  \includegraphics[width=8cm]{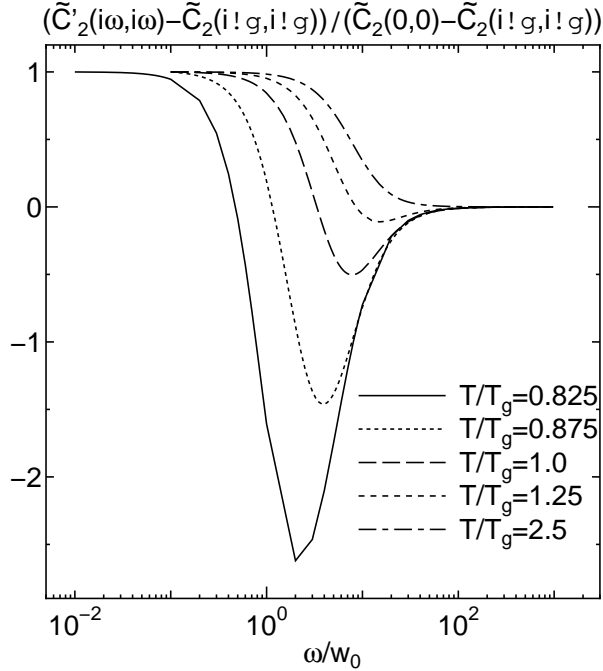}
 \end{center}
 \caption{The real part of the scaled 2nd-order ac specific heat
 at temperatures $T/T_g=$0.825, 0.875, 1.0, 1.25 and 2.5:
 As the temperature is decreased below the temperature $T_X/T_g = 1.25$,
a local minimum decreases.}
 \label{fig:real}
\end{figure}

\begin{figure}
 \begin{center}
  \includegraphics[width=8cm]{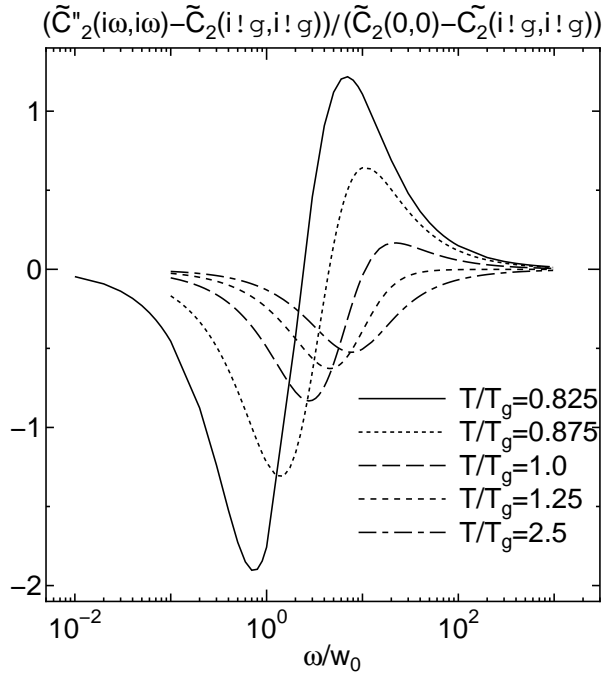}
 \end{center}
 \caption{The imaginary part of the scaled
 2nd-order ac specific heat
 at temperatures $T/T_g=$0.825, 0.875, 1.0, 1.25 and 2.5:
 As the temperature is decreased below the temperature $T_X/T_g = 1.25$,
 a local minimum decreases and a local maximum increases.}
 \label{fig:imag}
\end{figure}

The real and imaginary parts of 
the scaled 2nd-order ac specific heat
$(\tilde{C}_2(i\omega,i\omega)-\tilde{C}_2(i\infty,i\infty))/
(\tilde{C}_2(0,0)-\tilde{C}_2(i\infty,i\infty))$
are shown in Fig. \ref{fig:real} and Fig. \ref{fig:imag}, respectively.
The real part shows a decreasing local minimum 
when the temperature is reduced below the crossover temperature 
$T_X/T_g = 1.25$.
It is important to note that the similar behavior is shown for the
imaginary part in Fig.\ref{fig:imag},
where the imaginary part has a decreasing minimum and an
increasing maximum below the crossover temperature.

\begin{figure}
 \begin{center}
  \includegraphics[width=8cm]{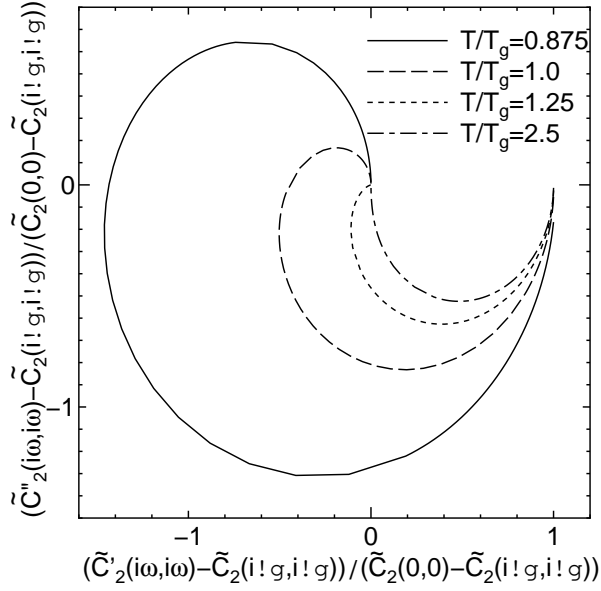}
 \end{center}
 \caption{The Cole-Cole plot of the scaled 2nd-order ac specific heat 
 at temperatures $T/T_g=$0.875, 1.0, 1.25 and 2.5:
 At a high temperature, the plots are semicircle.
 As the temperature is decreased, the plots deviate from the semicircle.}
 \label{fig:Trap_2nd_Cole}
\end{figure}

The deviation from the Debye relaxation appears clearly in the Cole-Cole plot
of the 2nd-order ac specific heat shown in Fig.\ref{fig:Trap_2nd_Cole}.
In the high temperature region, the plots show the semicircle,
since the 2nd-order ac specific heat is equal to the susceptibility of the
Debye relaxation.
As the temperature is reduced, the plots deviate from the semicircle,
that shows that the terms including $\partial W/\partial T$ in
Eqs. (\ref{eq:FELP_2nd_Real}) and (\ref{eq:FELP_2nd_Imag})
give significant contributions.

\begin{figure}
 \begin{center}
 \includegraphics[width=8cm]{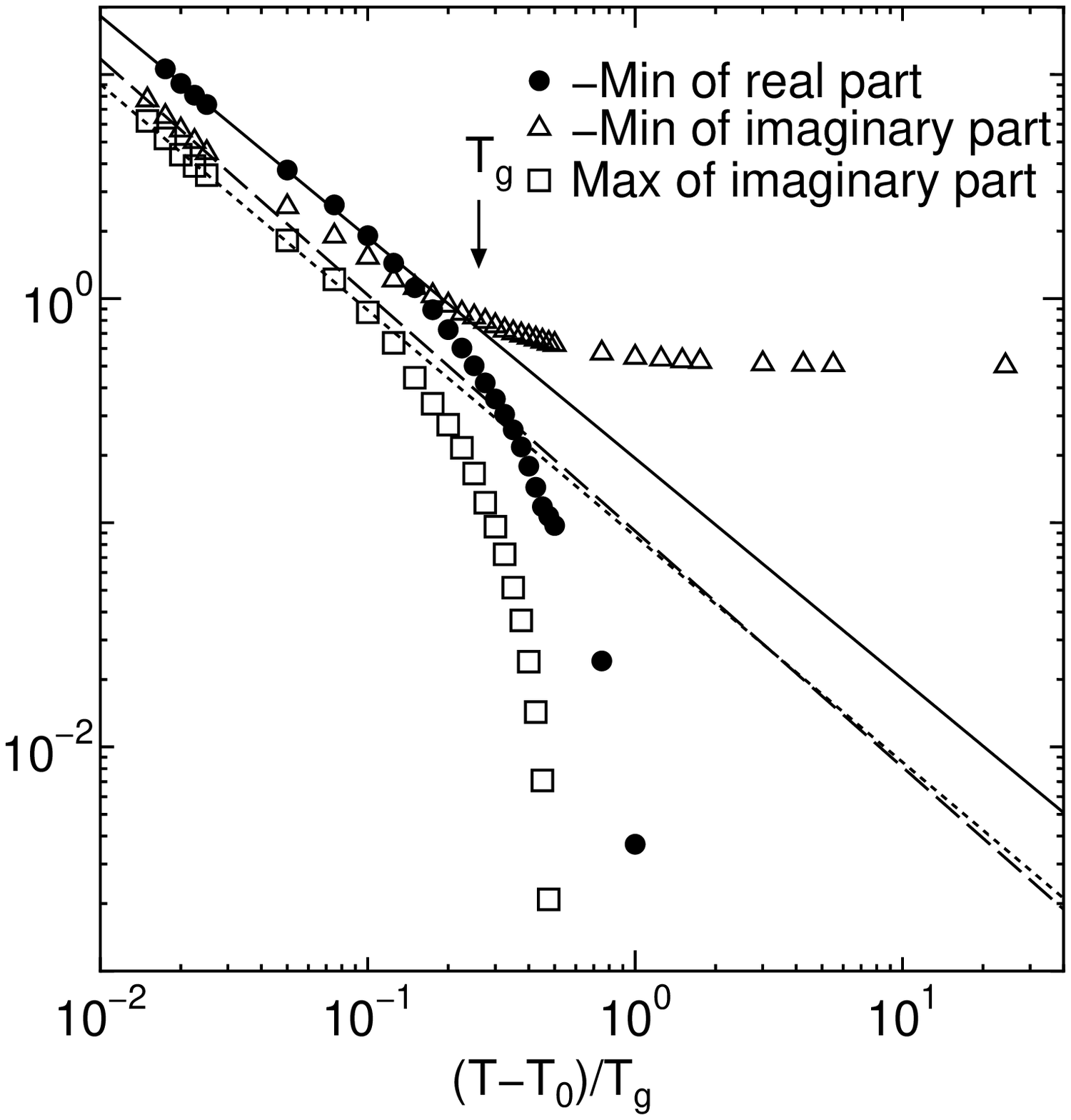}
 \end{center}
 \caption{The temperature dependence of the extrema of 
 the scaled 2nd-order ac specific heat:
 The lines are fitted by power law functions in the low
 temperature region.
 Below $T_g$, the extrema obeys the power law as $(T-T_0)^{-\nu}$.
 Above $T_g$, the temperature dependence of the extrema shows significant
changes.
 }
 \label{fig:extrema}
\end{figure}

The temperature dependence of the extrema of the 2nd-order ac specific
heat
is shown in Fig.\ref{fig:extrema}.
Below $T_g$, the extrema obey the power law as $(T-T_0)^{-\nu}$,
as the temperature is decreased toward $T_0$.
The exponents $\nu$ are $\nu=0.97, 1.05, 1.01$ for 
the local maximum of the real part, the local minimum and the local
maximum of the imaginary part, respectively.
Above $T_g$, the temperature dependence of the minimum of
the imaginary part deviates from the power law and approaches 
to a constant value.

\section{Conclusion}

We presented the theoretical formalism of calculating
the energy response to the oscillating temperature
in the free energy landscape picture.
The characteristic behaviors of the energy responses are summarized
as follows.

The large and small frequency limits of the 1st-order ac
specific heat are identical to the equilibrium and quenched specific heat in
the free energy landscape picture, respectively.
The distribution of the eigenvalues of the transition rate matrix
among basins is shown to give rise to the broad peak
of the imaginary part of the 1st-order ac specific heat.
This is consistent with the experiment for the glass forming materials.

The 2nd-order ac specific heat is defined for the first time
as the coefficient of
the oscillating part of the 2nd-order energy response.
The large and small frequency limits of the 2nd-order ac specific heat
are identical to the temperature derivative of the annealed and quenched
specific heat, respectively.

We analyzed the 1st- and 2nd-order ac specific heats for a model glass former
on the basis of the free energy landscape picture and
showed characteristics of the ac specific heats.

A frequency $\omega_{peak}$ at a peak of an imaginary part 
of the 1st-order ac specific heat
shifts to the low frequency region as the temperature is decreased 
below the glass temperature temperature $T_g$ 
and approaches to $0$ at the Vogel-Fulcher temperature $T_0$.
In the high temperature region above $T_g$,
$\omega_{peak}$ becomes virtually independent of the temperature
and approaches a constant value.
A width of the peak becomes larger than that of the
Debye relaxation type below $T_g$ due to the distribution of the eigenvalue
$\lambda$ and diverges at $T_0$.
The statistics of $\lambda$ such as $\langle\lambda\rangle$, $\langle\lambda^2\rangle$,
and $\langle\lambda^{-1}\rangle$ can be obtained from the frequency dependence
of the 1st-order ac specific heat, and these quantities
shows deviation from the Debye-type relaxation below $T_g$.

The scaled 2nd-order ac specific heats have extrema,
which obey the power law as $(T-T_0)^{\nu}$ below $T_g$.
Above $T_g$, the temperature dependence of the extrema changes.
The local minimum of the real part and the local maximum of the imaginary part
vanish and the local minimum of the imaginary part
approaches to the constant value near the crossover temperature $T_x$.

In conclusion, the characteristics of the transition rate among basins
reflect the frequency dependence of the 1st- and 2nd-order ac specific heats.
It means that the glass transition point $T_g$
, the Vogel-Fulcher temperature $T_0$
and the crossover temperature $T_x$
can be determined from the 1st- and 2nd-order ac specific heats.
It is interesting to note that we can determine $T_0$,
at which the shear viscosity diverges,
i.e. a dynamically defined characteristic temperature,
from the thermodynamic measurement as the 1st- and 2nd-order energy response.

\section*{Acknowledgments}

We would like to thank Professors Y. Saruyama, O. Yamamuro
and A. Yoshimori for useful discussions on this study.
This work was supported in part by a Grand-in-Aid for Scientific
Research from the Ministry of Education, Culture, Sports, Science and
Technology.

\end{document}